# Synchrotron radiation in the presence of a perfect cylindrical mirror.

Note on the interference of the direct radiation and the high-order whispering–gallery modes

**Sándor Varró**

Research Institute for Solid State Physics and Optics of the Hungarian Academy of Sciences

H-1525 Budapest, PO Box 49, Hungary, E-mail: varro@mail.kfki.hu

**Abstract.** Exact solutions are given for the Maxwell equations driven by a gyrating ultrarelativistic electron inside of a reflecting cylindrical boundary. The axis of the electron's circular trajectory and the axis of the surrounding cylinder are supposed to coincide. It is found that, at certain values of the ratio of the cylinder's and of the trajectory's radii, the field amplitudes diverge as functions of time. The physical reason for this is the constructive interference between the radiation emitted earlier and fed back by reflection to the actual position of the radiating electron.

## 1. Introduction

The cyclotron and synchrotron radiation in free space has long been thoroughly studied both theoretically and experimentally [1, 2, 3, 4]. Recently in the context of cyclotron masers [5, 6] and cavity electrodynamics, cavity effects have also been studied [7, 8]. In the cyclotron maser the fundamental or lowharmonics are generated, on the other hand, in the synchrotron radiation very high harmonics are dominant. The question naturally emerges whether could one somehow feed back these high harmonics in order to have stimulated emission and manage perhaps lasing also. In the present paper we give a short study of one possible feedback mechanism, where the trajectories of the gyrating ultrarelativistic electrons are completely surrounded by a coaxial mirror. We will show that, if the ration of the the cylinder's radius and the radius of the electron's trajectory sarisfies certain geometrical resonance conditions, then practically for all very high harmonics resonsnce can be achieved. We do not know yet whether our scheeme can be of relevance for a synchrotron laser, but we think that the problem discussed here is interesting itself, too, so that is worth presenting it.

## 2. Excitation of cylindrical wave-guide modes by an ultrarelativistic electron

In the present section we give an exact solution of Maxwell's equations driven by an ultrarelativistic electron gyrating inside of a perfectly reflecting cylindrical wave-guide.

The transverse position of the electron moving in the homogeneous magnetic field $\vec{e}_z B_0$ has the Cartesian components

$$x(t) = r_0 \cos(\omega_0 t + \varphi_0), \quad y(t) = r_0 \sin(\omega_0 t + \varphi_0), \quad z(t) = 0, \tag{1}$$

where $r_0 = \upsilon/\omega_0$, $\omega_0 = \omega_c/\gamma$ with $\omega_c = |e|B_0/mc$ being the cyclotron frequency, $\upsilon$ denotes the electron's velocity and $\gamma \equiv (1-\beta^2)^{-1/2}$, $\beta \equiv \upsilon/c$. We assume that the the longitudinal component of the electron's velocity is zero, i.e. $\upsilon_z = 0$, and the gyration takes place in the $z = 0$ plane. The charge density $\rho$ and the current density $\vec{j}$, associated to the trajectory of one single electron with initial phase $\varphi_0$ given by Eq. (1), can be conveniently expressed in cylindrical coordinates

$$\{\rho(r,\varphi,z;t),\ \vec{j}(r,\varphi,z;t)\} = \{1,\ \upsilon\vec{e}_\varphi\}er^{-1}\delta(r-r_0)\delta[\varphi-(\omega_0 t+\varphi_0)]\delta(z)u(t), \qquad (2)$$

where we have also introduced the unit step function $u(t)$, describing a sharp switching–on of the interaction. Equations (1) and (2) correspond to a highly idealistic situation for many reasons. Because in reality no perfectly homogeneous and stationary magnetic fields can be sustained over large spatio-temporal regions, exactly planar and circular trajectories of charges can never be secured, and, moreover the position of the guiding center cannot be fixed up to an arbitrary accuracy. The production and injection of a perfectly monoenergetic electron beam is impossible, either. The radiation loss necessarily distorts the trajectory, and the reacceleration cannot be solved without causing additional oscillations. In short, the spatio-temporal inhomogenities and spectral imperfections of both the boundary conditions and of the charged particle beams, of course, do not allow to prepare and sustain such an ideal current distribution shown in Eq. (2). The 'fragility' of this external current density may be a subject of a separate study, like the investigation of the possible bunching effect caused by the feedback of the radiation which would result in superradiance and lasing. In the present paper we shall not consider these important questions, rather we concentrate on the simplest, but exactly solvable part of the radiation problem.

In order to obtain a physically meaningful solutions of Maxwell's equations driven by the densities given by Eq. (2) inside the cylinder, we have to take into account the boundary conditions $[\vec{n}\times\vec{E}]_C = 0$ and $[\vec{n}\cdot\vec{B}]_C = 0$, where the subscript $C$ symbolizes the boundary of the cylinder of radius $a$ (larger than $r_0$), and $\vec{n}$ is the outward unit normal of it. In words this means that the tangential component of the electric field strength $\vec{E}$ and normal component of the magnetic induction $\vec{B}$ must vanish at the surface of the cylinder (at any vertical position $z$). Thus $\vec{E}$ and $\vec{B}$ are expanded into a superposition of the so-called cross-sectional vector eigenfunctions,

$$\vec{E} = \sum_{m\,p} a_{mp}\vec{\nabla}_\perp\Phi_{mp} + \sum_{n\,s} b_{ns}\vec{e}_z\times\vec{\nabla}_\perp\Psi_{ns} + \vec{e}_z\sum_{m\,p} c_{mp}\Phi_{mp}, \qquad (3a)$$

$$\vec{B} = \sum_{m\,p} \alpha_{mp}\vec{e}_z\times\vec{\nabla}_\perp\Phi_{mp} + \sum_{n\,s} \beta_{ns}\vec{\nabla}_\perp\Psi_{ns} + \vec{e}_z\sum_{n\,s}\gamma_{ns}\Psi_{ns}. \qquad (3b)$$

In Eqs. (3a,b) $\Phi_{mp}$ and $\Psi_{ns}$ are Dirichlet and Neumann eigenfunctions satisfying the scalar Helmholtz equation $(\nabla_\perp^2 + k^2)f = 0$ with eigenvalues $k_{mp}$ and $k_{ns}$, respectively. The unknown coefficients $a_{mp}$, $b_{ns}$, $c_{mp}$ and $\alpha_{mp}$, $\beta_{ns}$, $\gamma_{ns}$ are to be determined as functions of time ($t$) and vertical position ($z$). According to the boundary condition $[\Phi_{mp}]_C = 0$, $\Phi_{mp}$ must have the form

$$\Phi_{mp} = J_m(x_{mp}(r/a)) \begin{Bmatrix} \sin(m\varphi) \\ \cos(m\varphi) \end{Bmatrix}, \quad J_m(x_{mp}) = 0, \tag{4a}$$

where $x_{mp}$ is the $p$-th root of the ordinary Bessel function of first kind $J_m$ of order $m$. Because of the Neumann boundary condition, $[\partial \Psi_{ns}/\partial r]_C = 0$, we have

$$\Psi_{ns} = J_n(y_{ns}(r/a)) \begin{Bmatrix} \sin(n\varphi) \\ \cos(n\varphi) \end{Bmatrix}, \quad J'_n(y_{ns}) = 0, \tag{4b}$$

where $y_{ns}$ is the $s$-th root of the derivative $J'_n$ of the the ordinary Bessel function of first kind of order $n$. The eigenvalues of the corresponding wave numbers are $k_{mp} = x_{mp}/a$ and $k_{ns} = y_{ns}/a$, respectively, where $a$ is the radius of the cylinder. By taking the into account the orthogonality property of the cross-sectional eigenfunctions, we can derive from the inhomogeneous Maxwell equations two sets of coupled first order differential equations for the expansion coefficients. The set $\{a_{mp}, \alpha_{mp}, c_{mp}\}$ is responsible for the dynamics of the TM and longitudinal components of the electromagnetic field. On the other hand, the dynamics of the TE components is governed by the set $\{b_{ns}, \beta_{ns}, \gamma_{ns}\}$. It can be shown that $\alpha_{mp}(z=0,t) = 0$ and $c_{mp}(z=0,t) = 0$, moreover, in the case we shall discuss below $a_{mp}$ vanishes to a good approximation at any vertical position. This means that in the plane of the electron gyration only the TE modes are excited, and that is why henceforth we shall be dealing with only the behaviour of the TE modes.

The coupled system of equations for $b_{ns}$, $\beta_{ns}$ and $\gamma_{ns}$ reads

$$\frac{\partial \beta_{ns}}{\partial z} - \frac{1}{c}\frac{\partial b_{ns}}{\partial t} - \gamma_{ns} = \frac{4\pi}{c}\frac{1}{N_{ns}^2}\int d^2s \vec{j} \cdot (\vec{e}_z \times \vec{\nabla}_\perp \Psi_{ns}), \tag{5a}$$

$$\frac{\partial b_{ns}}{\partial z} - \frac{1}{c}\frac{\partial \beta_{ns}}{\partial t} = 0, \tag{5b}$$

$$b_{ns} - \frac{1}{ck_{ns}^2}\frac{\partial \gamma_{ns}}{\partial t} = 0, \tag{5c}$$

where $N_{ns}^2 = (\pi/\varepsilon_n)J_n^2(y_{ns})(y_{ns}^2 - n^2)$, and $\varepsilon_1 = 1$ and $\varepsilon_n = 2$ for $n = 2,3,....$ The integration on the rhs of Eq. (5a) is to be evaluated over the cross-section of the cylinder. Having eliminated the functions $\beta_{ns}$ and $\gamma_{ns}$ from Eqs. (5a,b,c) we arrive at an inhomogeneous Klein-Gordon equation for $b_{ns}(z,t)$,

$$\left(\frac{\partial^2}{\partial z^2} - \frac{1}{c^2}\frac{\partial^2}{\partial t^2} - k_{ns}^2\right)b_{ns} = B_{ns}\delta(z)f'_n(t)/c, \tag{6a}$$

where

$$B_{ns} \equiv 4e\beta \frac{(y_{ns}/a)J'_n(y_{ns}r_0/a)\varepsilon_n}{J_n^2(y_{ns})(y_{ns}^2 - n^2)}, \quad f_n(t) \equiv \begin{Bmatrix} \sin[n(\omega_0 t + \varphi_0)] \\ \cos[n(\omega_0 t + \varphi_0)] \end{Bmatrix} u(t). \tag{6b}$$

The Green's function of Eq. (6a) can be derived by using standard methods

$$g(t,z) = -\frac{c}{2} J_0\left(ck_{ns}\sqrt{t^2 - z^2/c^2}\right) u\left(t - \frac{|z|}{c}\right), \quad t \equiv t_1 - t_2, \quad z \equiv z_1 - z_2. \tag{7}$$

With the help of the Green's function in Eq. (7), the solution of Eq. (6a) can be determined by straighforward integrations. The general explicit form of $b_{ns}$ is a complicated expression, however for large values of $t$ a relatively simple expression can be derived (for short, we present only the upper component stemming from the sine oscillations):

$$-2b_{ns}(z,t)/B_{ns} = J_0\left(\omega_{ns}\sqrt{t^2 - z^2/c^2}\right)(\sin \varphi_0) u\left(t - \frac{1}{c}|z|\right)$$

$$+ \frac{\nu_{ns} u(\nu_{ns} - 1)}{\sqrt{\nu_{ns}^2 - 1}} \sin n\left[\omega_0\left(t - \frac{1}{c}\sqrt{\nu_{ns}^2 - 1}|z|\right) + \varphi_0\right] \tag{8}$$

$$+ \frac{\nu_{ns} u(1 - \nu_{ns})}{\sqrt{1 - \nu_{ns}^2}} (\cos n(\omega_0 t + \varphi_0)) \exp\left[-\frac{n\omega_0}{c}\sqrt{1 - \nu_{ns}^2}|z|\right]$$

where $\nu_{ns} \equiv n\omega_0/\omega_{ns}$ with $\omega_{ns} \equiv ck_{ns}$ being the TE eigenfrequencies. The first term on the rhs of Eq. (8) represents a (transient) *precursor* with front velocity $c$, which vanishes as $1/\sqrt{t}$. The second and the third terms correspond to *above–cutoff* and *below–cutoff waves*, respectively, the latter being bound to the $z = 0$ plane. At the exact resonances $\nu_{ns} = 1$, Eq. (8) loses its validity, and $b_{ns}$ has a qualitatively different form (for shortwe present only the upper component of $b_{ns}$ taken at the $z = 0$ plane (of the electron's trajectory):

$$-2b_{ns}/B_{ns} = J_0(n\omega_0 t)(\sin n\varphi_0) + (n\omega_0 t)[J_0(n\omega_0 t)(\cos n\varphi_0) - J_1(n\omega_0 t)(\sin n\varphi_0)]. \tag{9}$$

On the basis of the asymptotic behaviour of the Bessel functions, it can be shown that for large $t$ times $b_{ns}$ diverge as $\sqrt{t} \times$(oscillatory function). Of course, since some sort of damping is always present in physical systems, such a divergence is not realistic. If we introduce the phenomenological damping term $(-k_{ns}/cQ_{ns})(\partial b_{ns}/\partial t)$ into the Klein-Gordon equation on the lhs of Eq. (6), then the solution has a similar structure as that shown in Eq. (8), but with the essential difference that the potential unphysical divergences are replaced by finite resonance terms. The oscillatory parts will contain resonance denominators of the form $[(\nu_{ns}^2 - 1)^2 + \nu_{ns}^2/Q_{ns}^2]^{1/4}$, where $Q_{ns}$ is spectral Q-factor related to the assumed finite conductivity of the cylinder wall. Thus, close to resonance the amplitudes are increased by a factor of $\sqrt{Q_{ns}}$. For the high harmonics, we are interested in, $Q_{ns}$ can be well approximated by $a/\delta_{ns}$, where $\delta_{ns} = (2/\mu\sigma\omega_{ns})^{1/2}$ is the skin depth for a spectral component. For example, for silver $\delta_\omega \approx 6 \times 10^{-6}\, cm$ for $\omega/2\pi \approx 10^{10}\, Hz$. In the optical region $\delta_\omega$ can well be of two orders of magnitude smaller, thus $Q_\omega$ can be very large if $a$ is of order of meters, say.

## 3. Resonance conditions

In the present section we study the question of under what conditions simultaneous resonance can be reached for the 'most of the higher harmonics' in the synchrotron radiation in the cylindrical mirror.

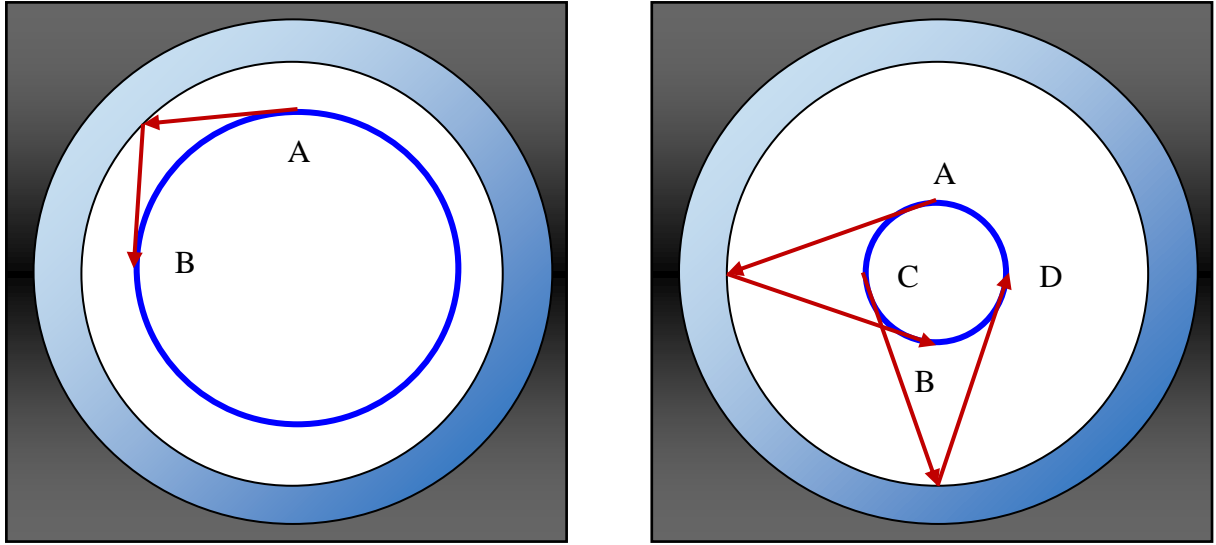

**Fig. 1.** A ray of radiation emanating tangentially within a narrow cone from the electron at point $A$, gets reflected on the cylindrical mirror and arrives at point $B$ exacly at the time when the electron arrives there. Left: illustrates the zeroth geometrical resonance condition (N=0). Right: illustrates the first resonance condition (N=1). In the latter case, after one complete revolution the electron encounters with its own radiation field (which was emitted tangentially one turn earlier). For further explanation see text in section 3.

The geometrical arrangement we are interested in is shown in Fig. 1. On *geometrical resonance condition* we mean that the radii of the electon's trajectory and of the cylinder are adjusted in such a way, that a signal emanating tangentially at point $A$, after reflection, gets back to the electron's trajectory at point $B$ exactly at that time when the electron (possibly after $N \geq 1$ complete revolutions, or even 'almost immediately') gets to the same point $B$. It is clear that, if once this condition is satisfied for the pair of points $A$ and $B$, then it will be satisfied for the pair $C$ and $D$, which can be obtained by rotating the pair $A$ and $B$ by an arbitrary angle. *In this way the electron 'after a while' will continuously move 'in phase' in its own retarded radiation field which has been emitted earlier at different points on the trajectory.* We think that this arrangement would secure an effective feedback for obtaining stimulated emission. Needless to say, the correctness of the expectation suggested by this intuitive picture should be checked on the basis of the accurate analytic treatment picture (which was originally just the starting point of the present investigation). By simple kinematic considerations it can be shown that, if the geometrical resonance condition holds, then the ration $a/r_0$ satisfies the following transcendental equation

$$\beta\sqrt{x^2-1} - \arccos(1/x) = N\pi, \quad x \equiv a/r_0, \tag{10}$$

where $N$ is the number of complete revolutions of the electron before the first ecounter with it own radiation field after one reflection. For $N=1$ we obtain approximately $a/r_0 \cong 3\pi/2$, and for $N=0$ we have $a/r_0 \cong 1 + 3/4\gamma^2$. It is clear that for an ultrarelativistic electron the latter 'zeroth resonant condition' can be satisfied if $a$ is only slightly larger than the radius of the trajectory, i.e. the electron moves very close to the inner surface of the cylinder. If we assume $r_0 \approx 10 cm$ in a strong confining magnetic field, then the the first resonance condition can be satisfied in a cylinder of radius $a \approx 47 cm$. Henceforth we shall study the case $N=1$.

The wave resonance condition $1 = \nu_{ns} \equiv n\omega_0/\omega_{ns} = (n/y_{ns})(\beta a/r_0)$, on the basis of the previous section, can be studied by using the asymptotic form of the zeros $y_{ns}$ of the derivatives $J'_n$ of the Bessel functions $J_n$. Here we restrict the discussion to the case when not only $n$ but also the $s$ are large. (It can be shown that if $s$ considerably differs from $n$, then the resonance cannot be reached.) For large $n$ we have

$$\frac{a}{r_0} \cong \beta\frac{a}{r_0} = \frac{y_{ns}}{n} = z(\zeta) + O\left(\frac{1}{n^2}\right), \quad \sqrt{z^2-1} - \arccos(1/x) = \frac{2}{3}(-\zeta)^{3/2}, \quad \zeta = n^{-3/2}a'_s. \tag{11}$$

The function $z(\zeta)$ is the inverse function defined by the second equation of Eq. (11), with $a'_s$ being the $s$-th negative zeros of the Airy function. Since $n$ is very large , $z = x$, and because $\beta = \upsilon/c$ is practically unity, the left hand sides of Eq. (10) coincides with the left hand side of the second equation of Eq. (11). Now, for large $s$ values the zeros can be approximated by the analytic formula $-\zeta = (3\pi s/2n)^{2/3}$. By having taken this relation into account, we can easily check that, if the geometrical resonance condition is secured, then the wave resonance condition in Eq. (11) becomes an identity for $s = n$. As a consequence, *the wave resonance condition is independent of the $n$-values, provided these are large enough*. This means that (if $Q_{ns}$ is a smooth function of $n$) there exist a broad band of the spectrum which is uniformly 'lifted-up' due to these resonances. *This resonant accumulation process can be interpreted as a result of the constructive interference between the radiation actually emitted and the self- radiation emitted earlier and fed back (by reflection) to the actual position of the electron.*

## 4. Summary

In the present paper we have discussed some of the characteristics of the synchrotron radiation emitted by a single ultrarelativistic electron in the interior of a coaxial cylindrical mirror. The combined initial-value and boundary-value problem of the Maxwell equations have been solved exactly, but neither the radiation reaction, nor the role of spatio-temporal imperfections have been discussed. It was shown that near the plane of the electron's gyration mostly the TE modes are excited, and the effect of damping at resonance has also been briefly discussed. In section 3 we have introduced the concept of *geometrical resonance*, whose notion is based on a simple ray construction. It was shown that if this condition is satisfied, then the exact *wave resonance condition* does not depend on the excitation indeces of the very high harmonics. At such a 'broad-band resonance' there is an accumulation process taking place, due to which the intensity of the resulting radiation can be increased by orders of magnitudes. The present analysis supports our original physical picture according to which, if the electron meets with its own radiation field emitted earlier in a tangential narrow cone, and reflected back by the mirror, then there is a constructive interference between this retarded self-field and the actually emitted radiation field.

**Acknowledgement.** This work has been supported by the Hungarian National Scientific Research Foundation OTKA, Grant No. K73728.